\documentclass{iopart}
\usepackage{iopams,cite}  
\usepackage{graphicx}
\begin{document}

\title[Gyromagnetic ratio of relativistic stars]{Gyromagnetic ratio of
rapidly rotating compact stars in general relativity} 

\author{J\'er\^ome Novak\dag\ 
\footnote[3]{To whom correspondence should be addressed
(Jerome.Novak@obspm.fr)} 
and Emmanuel Marcq\dag\ddag\ }

\address{\dag\ Laboratoire de l'Univers et de ses Th\'eories (UMR8102
du C.N.R.S.), Observatoire de Paris -- Section de Meudon, F-92195 Meudon
Cedex, France.}

\address{\ddag\ \'Ecole Normale Sup\'erieure, 45 rue d'Ulm, F-75230
Paris Cedex 05, France.}

\begin{abstract} 
We numerically calculate equilibrium configurations of uniformly rotating
and charged neutron stars, in the case of insulating material and
neglecting the electromagnetic forces acting on the equilibrium of the
fluid. This allows us to study the behaviour of the gyromagnetic ratio
for those objects, when varying rotation rate and equation of state
for the matter. Under the assumption of low charge and incompressible
fluid, we find that the gyromagnetic ratio is directly proportional to
the compaction parameter $M/R$ of the star, and very little dependent
on its angular velocity. Nevertheless, it seems impossible to have
$g=2$ for these models with low charge-to-mass ratio, where matter
consists of a perfect fluid and where the collapse limit is never
reached.  
\end{abstract}


\submitto{\CQG}


\section{Introduction}\label{s:intro}
For steadily rotating, charged massive bodies, the gyromagnetic factor $g$ is
defined as the ratio
\begin{equation}
g = \frac{2M{\cal M}}{QJ}; \label{e:def}
\end{equation}
where $M$ is the mass, $Q$ the total charge and $J$ the angular
momentum of the system. $\cal M$ is the magnetic moment of the system,
linked with the motion of the charges that are accounting for
$Q$. With such a definition, in 
classical electrodynamics a rotating charged particle has a
gyromagnetic factor equal to 1. The same is thus true for any system
in classical electrodynamics, with a constant ratio of charge and mass
density. On the other hand, within general relativity, the
gyromagnetic ratio for all charged and rotating black holes is
$g=2$. Many points concerning the gyromagnetic ratio for isolated
systems within classical electrodynamics, quantum theory and general
relativity have been studied by Pfister and King \cite{PfiK03}. The
question we try to address here is the following one. How does the
gyromagnetic factor behave for ``intermediate'' objects in general
relativity, that possess a
gravitational field weaker than that of black holes, but in which
strong field effects are not negligible?

The aim of this paper is to try to answer
this question by numerically studying the $g$-factor of rotating and
charged relativistic compact stars, within the framework of general
relativity. We use a self-consistent physical model, in which matter
is supposed to be a charged (insulator type) perfect fluid;
axisymmetric and in stationary rotation. As it will be shown later,
the only limitation shall be on neglecting the electromagnetic forces
acting on hydrodynamic equilibrium, making
our study a ``low charge'' approximate one; but we take into
account the electromagnetic field contribution to the total
energy-momentum tensor. The variation of the
gyromagnetic ratio as a function of the number of particles in the
system, or as a function of the rotation frequency will also be
discussed. Details of the physical model are presented in
section~\ref{s:model}, section~\ref{s:resu} is then giving results of the
numerical study for two equations of state, which determine local properties of
matter, and making comparison with previous works. Finally,
section~\ref{s:conc} summarises results and gives some concluding remarks.

\section{Model and assumptions}\label{s:model}
We here give the most important assumptions made in our study;
complete details of the formalism and the way numerical stars are
computed can be found in Bonazzola {\it et al.\/} \cite{BonGSM93} and
in Bocquet {\it et al.\/} \cite{BocBGN95} for the magnetised configurations.

\subsection{Stationary and axisymmetric spacetime}\label{ss:field}

We want to solve the coupled Einstein-Maxwell equations to get general
relativistic magnetised models of stationary rotating bodies. We make
the assumption that the spacetime is also stationary (asymptotically
timelike Killing vector field), axisymmetric (spacelike Killing vector
field which vanishes on a timelike 2-surface, {\em axis of symmetry}
and whose orbits are closed curves) and asymptotically flat. In
addition, we suppose that the source of the gravitational field
satisfies the {\em circularity condition}, equivalent to the absence
of meridional convective currents and only {\em poloidal} magnetic
field is allowed. We then use MSQI (Maximal Slicing - Quasi Isotropic, see
\cite{BonGSM93}) coordinates $(t, r, \theta, \varphi)$, in which the
metric tensor takes the form
\begin{equation}
\fl \rmd s^2 = g_{\mu\nu}\rmd x^\mu \rmd x^\nu = -N^2 \rmd t^2 +
B^2r^2\sin^2 \theta (\rmd \varphi -
\beta^\varphi \rmd t)^2 + A^2(\rmd r^2 + r^2 \rmd \theta^2), \label{e:defmet}
\end{equation}
where $N, \beta^\varphi, A$ and $B$ are four functions of
$(r,\theta)$. 

With our hypothesis, the electromagnetic field tensor $F_{\alpha
\beta}$ must be derived from a potential 1-form with the following
components
\begin{equation}
A_\alpha = (A_t, 0, 0, A_\varphi). \label{e:defamu}
\end{equation}
The Einstein-Maxwell equations result in a set of six coupled
non-linear elliptic equations for the four metric and the two
electromagnetic potentials (see \cite{BonGSM93} and
\cite{BocBGN95}). The right-hand side of this system also involves
matter terms (density and charge currents), which will be discussed in
next section. 

\subsection{Fluid properties}\label{ss:fluid}

The matter is supposed to consist of a perfect fluid, so there exists
a privileged vector field: the 4-velocity $u^\alpha$. The
energy-momentum tensor takes its usual form
\begin{equation}
T^{\mu\nu} = (e+p)u^\mu u^\nu + p g^{\mu\nu} + T^{\mu\nu}_{\rm EM},
\label{e:tmunu} 
\end{equation}
where $p$ is the fluid pressure and $e$ the energy density measured in
the fluid frame. $T^{\mu\nu}_{\rm EM} = 1/(4\pi) \left( F^{\mu \alpha}
F^\nu_\alpha - 1/4 F_{\alpha\beta}F^{\alpha\beta} g^{\mu\nu} \right)$
is the electromagnetic contribution to the energy-momentum
tensor. Following \cite{BonGSM93}, we note $\Omega = 
u^\varphi / u^t$ and define $\Gamma$ as the Lorentz factor linking the
fluid comoving observer and the locally non-rotating one. We make the
assumption that the matter is rigidly rotating ($\Omega =$constant)
and that the equation of state (EOS) is a one parameter 
EOS (ignoring the influence of temperature): $e=e(n_{\rm B})$ and $p=p(n_{\rm B})$,
with $n_{\rm B}$ being the proper baryon density. 

The momentum-energy conservation gives an equation of stationary
motion for the fluid, which can be written as a first integral of
motion (the electromagnetic contribution to the energy-momentum tensor
is not considered yet), see \cite{BonGSM93}
\begin{equation}
\ln \frac{e+p}{n_{\rm B}} + \ln N -\ln \Gamma = \rm{constant}. \label{fim0}
\end{equation}
We turn now to the electromagnetic part; in order to have a complete
charged body, with a constant ratio of charge and mass density (see
\cite{PfiK03}), we suppose, contrary to \cite{BocBGN95} or
\cite{CarPL01}, that our 
system consists of an insulator, so that currents originate only from
macroscopic charge movement. The 4-current is thus proportional to
$j^\mu \sim u^\mu / \Gamma$ implying that $j^\varphi = \Omega
j^t$. Taking then electromagnetic force term $f_i = F_{i\sigma}j^\sigma/(e+p)$
into account, the momentum-energy conservation reads
\begin{equation}
\partial_i \left( \ln \frac{e+p}{n_{\rm B}} + \ln N -\ln \Gamma \right) +
\frac{1}{e+p} j^t\partial_i \left( \Omega A_\varphi - A_t \right) = 0.
\label{e:equi1} 
\end{equation}
The integrability condition of this equation is that the last term is
a gradient, so 
that there exists a function $M(r,\theta)$ such that
$j^t\partial_i \left( \Omega A_\varphi - A_t \right) = (e+p)\partial_i
M$. Following the same arguments as in \cite{BonGSM93}, there must exist
a regular function $m$, such that
\begin{equation}
j^t = (e+p)m\left( \Omega A_\varphi - A_t \right), \quad {\rm with } \quad
M(r,\theta) = \int_0^{\Omega A_\varphi(r,\theta) - A_t(r,\theta)} m(x)
\rmd x. \label{e:defgM}
\end{equation}

But, this gives too large a freedom for the distribution of
charged particles inside the star. In particular, the charge density $j^t$
is independent of the baryon density $n_{\rm B}$, and the $g$-factor
can, in principle, take any value. It is also irrelevant to compare
the $g$-factor obtained in this way, with its value 1 in classical
electrodynamics, where it is supposed that charge currents are
directly proportional to mass currents and charge density to mass
density \cite{PfiK03}. So we replace (\ref{e:defgM}) by
\begin{equation}
j^t = \chi n_{\rm B}, \quad {\rm and } \quad M=0, \label{e:prop}
\end{equation}
$\chi$ being the constant ratio between the charge and particle
densities, it is an input parameter (together with the central density
and the angular velocity) that controls the total charge of the system. 
It means that we do not integrate exactly momentum-energy
conservation equation (\ref{e:equi1}), since we neglect the
electromagnetic forces. It 
will be shown in section~\ref{s:resu} that this assumption is valid
for low total charges, where the electromagnetic forces are indeed negligible,
when compared to pressure, gravitational and centrifugal forces.

\subsection{Accuracy indicators}\label{ss:err}

\begin{figure}
\begin{center}
\vspace{6mm}

\includegraphics[height=8cm]{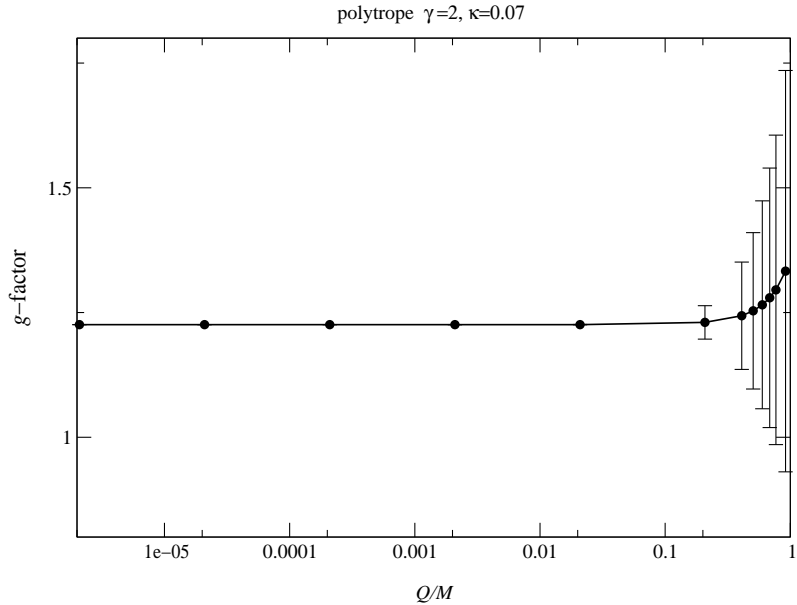}
\end{center}
\caption{Variation of the $g$-factor as a function of the
dimensionless ratio (total
charge) / (gravitational mass),
for a polytrope. Angular velocity $\Omega$ and baryon mass $M_{\rm B}$
are constant, being respectively equal to 200 Hz and $1.6
M_\odot$. Error bars are given by the GRV2 test.}\label{f:polyq} 
\end{figure}

To solve the six elliptical Poisson-like equations described in
section~\ref{ss:field}, we use {\em spectral methods} as described by
Grandcl\'ement {\it et al.\/} \cite{GraBGM01}. The complete numerical
procedure is presented in \cite{BocBGN95} as well as many tests of the
numerical code. Let us here emphasise that for our computations of
spacetimes, we have very reliable and
independent tests through the virial identities GRV2 (Bonazzola
\cite{Bona73}, Bonazzola and Gourgoulhon \cite{BonG94}) and GRV3
(Gourgoulhon and Bonazzola \cite{GouB94}), this latter being a
relativistic generalisation of the classical virial theorem. GRV2 and
GRV3 are integral identities which must be satisfied by any solution
of the Einstein-Maxwell equations we solve here, and are not imposed
during the numerical procedure. They are very sensitive to any
physical inaccuracy in the model, including eventual problems in the
equation of state. In the following, when presenting
accuracy of numerical results, we will refer to the accuracy by which
the numerical solution satisfies these virial identities. With the
exception of results shown in figure~\ref{f:polyq}, where error bars
are displayed, we only show results with better relative accuracy than
$10^{-5}$. 

As presented by Bonazzola {\it et al.\/} \cite{BonGSM93}, a key point
of the numerical method is to be able to integrate Einstein-Maxwell
equations up to spatial infinity, using a change of variable of the
type $u=1/r$ outside the star. This allows us to impose exact boundary
conditions at $r\to \infty$, and to compute global quantities from
asymptotic behaviour of the fields or integrals over the whole
space. We can therefore compute values of the total gravitational mass
$M$ and the total angular momentum $J$, from the gravitational field
$g_{\mu\nu}$; the total charge $Q$ and the magnetic moment $\cal M$
from the electromagnetic potential $A_\mu$. Another global quantity
characterising the star is the {\em circumferential radius} $R_{\rm
circ}$, defined as star's equatorial circumference (measured by the
metric (\ref{e:defmet})) divided by $2\pi$
\begin{equation}
R_{\rm circ} = B\left( R,\frac{\pi}{2} \right) R, \label{e:defrcirc}
\end{equation}
where $R$ is the coordinate equatorial radius. Finally, we shall use the
total baryon number of the star and its baryon mass $M_{\rm B}$.

\section{Numerical studies}\label{s:resu}

To calculate the $g$-factor of a given model, in addition to the
choice of a particular equation of state, one has to set the three
following parameters: the central density $n_{\rm B}(r=0)$ (or,
equivalently, the central log-enthalpy $\left. \ln (e+p)/n_{\rm B}
\right|_{r=0}$), the angular velocity $\Omega$ and the ratio between
mass and charge densities $\chi$ (\ref{e:prop}). 

\subsection{Polytropes}\label{ss:poly}

In this part we choose the EOS to be a polytropic one, of the form
(6.40) of reference \cite{BonGSM93}: $p=\kappa n_{\rm B}^\gamma$. We took
$\gamma = 2$ and $\kappa = 0.07 \rho_{\rm nuc}c^2/n_{\rm nuc}^2$,
where nuclear density $\rho_{\rm nuc} = 1.66 \times 10^{17} {\rm
kg/m^3}$ and $n_{\rm nuc} = 0.1 {\rm fm}^{-3}$. We first want to test
the validity of our assumption neglecting electromagnetic forces in the
equilibrium of the fluid. We computed a sequence of configurations
increasing the total charge, at fixed
angular velocity $\Omega =200$ Hz and fixed number of baryons
(equivalent to 1.6 solar mass). Results for the $g$-factor
(\ref{e:def}) as a function of the dimensionless ratio $Q/M$ are
displayed in figure~\ref{f:polyq}, together with the errors given by
the virial identities (see section~\ref{ss:err}). 
For $Q/M \lesssim 0.01$, $g$ is constant at $10^{-5}$ accuracy. For
$Q/M\gtrsim 0.01$, $g$ starts to vary, but this variation remains
within error bars, that become very important as $Q/M \to 1$. This
indicates that the fact that we are neglecting electromagnetic forces in the
equilibrium of the star induces an error that is lower than the
numerical one, as long as $Q/M \lesssim 0.01$. It may be seen as a
``low charge'' approximation for our model and, {\em within this
approximation}, we have checked with different equations of state
(other polytropes, incompressible fluid EOS, strange matter EOS) that
the $g$-factor would not depend on the charge. Let us say here that,
although we are neglecting electromagnetic forces in the fluid
equilibrium, we are taking the electromagnetic field into account in
the sources of Einstein equations.
\begin{figure}
\begin{center}
\vspace{6mm}

\includegraphics[height=8cm]{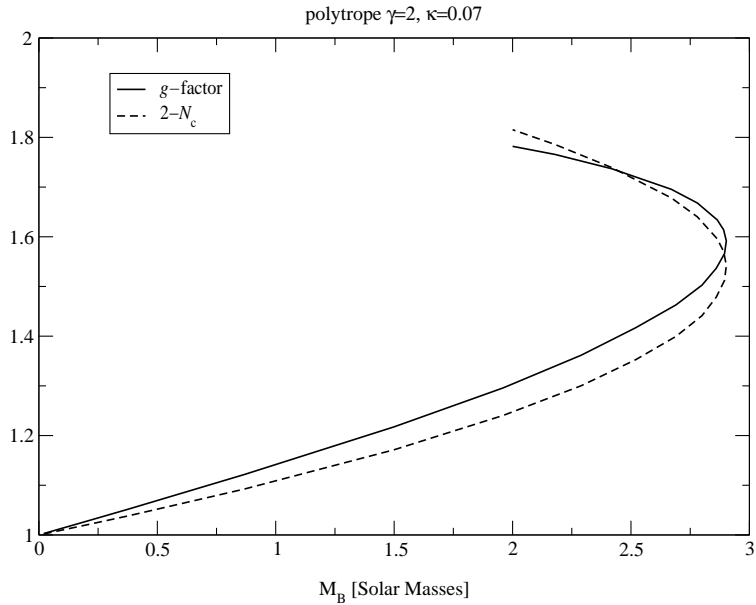}
\end{center}
\caption{Variation of the $g$-factor and of the central lapse
$N_c=N(r=0)$, as a function of baryon mass, for a
polytrope. Angular velocity $\Omega = 10$ Hz is kept
constant.}\label{f:polymbnu} 
\end{figure}
 
The regime for which $Q/M>1$ does not seem realistic for a perfect
fluid. Indeed, the repulsive Coulomb force acting on charged particles
becomes comparable to the gravitational force and pressure. Therefore,
there may not exist any stationary configuration for $Q/M$ too larger
than 1: the electrostatic force would overcome the gravitational one and
disperse particles. Moreover, Mustafa {\it et al.\/} \cite{MusCP87}
have shown that, in the case of a slowly rotating charged shell, when
$Q/M>1$ there is no upper or lower bound on the value of the
$g$-factor. Finally, in
order to find an acceptable stationary solution for $0.01 \lesssim Q/M
< 1$, one would have to satisfy both equations (\ref{e:defgM})
and $j^t = \chi n_{\rm B}$. These are, in general, incompatible for a
constant angular velocity $\Omega$ and one would have to 
allow for differential rotation of the
fluid. This has not been done in our study and it would certainly be
an improvement of our work. In the 
following, we will stay at $Q/M = 10^{-3}$ an will consider that $g$
is independent of the total charge, in the low charge regime.  

\begin{figure}
\begin{center}
\vspace{6mm}

\includegraphics[height=8cm]{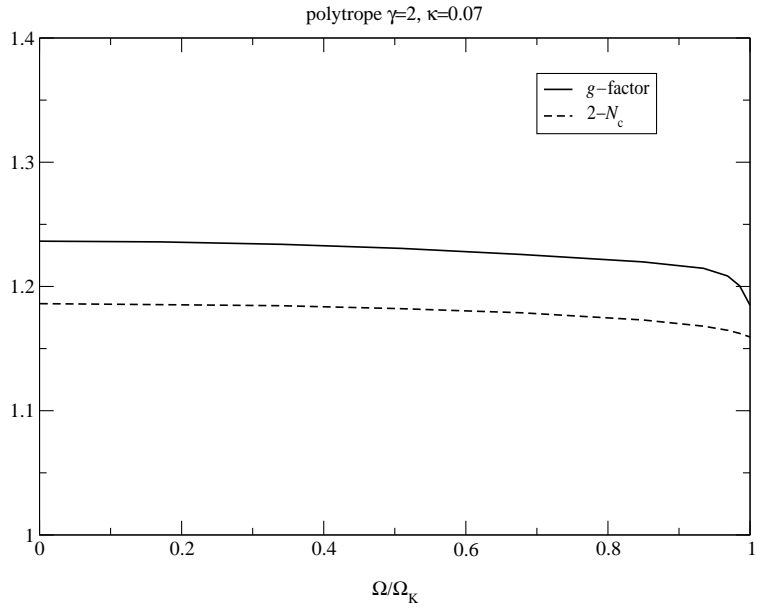}
\end{center}
\caption{Variation of the $g$-factor and of the central lapse
$N_c=N(r=0)$, as a function of the angular velocity, for a
polytrope. The baryon mass $M_{\rm B} = 1.6 M_\odot$ is kept
constant.}\label{f:polyfmb}  
\end{figure}

Now, we look at the variation of $g$, when the number of particles
(the baryon mass $M_{\rm B}$) of the rotating polytrope is changed. Results
for the $g$-factor are displayed in figure~\ref{f:polymbnu} (solid
line), together with the value of the lapse $N$ at the centre of the
star (dashed line). The parameter varying along 
both curves is the central density and we retrieve the well-known
result of the existence of a maximal mass for those stars. Thus, the
higher branch of each curve corresponds to unstable
configurations. The Newtonian limit $g=1$ is recovered at low baryon
masses, corresponding to weak gravitational field ($N_c \to 1$). In
more relativistic regime, the $g$-factor follows roughly the variation of the
central lapse, never reaching the value of 2 (corresponding to $N_c =
0$), which 
corresponds to a charged black hole. The maximal value that could be
reached (for an {\em unstable} configuration) was $g=1.8$, if only
{\em stable} solutions are considered, then one has $g\leq 1.6$. 

\begin{figure}
\begin{center}
\vspace{6mm}

\includegraphics[height=8cm]{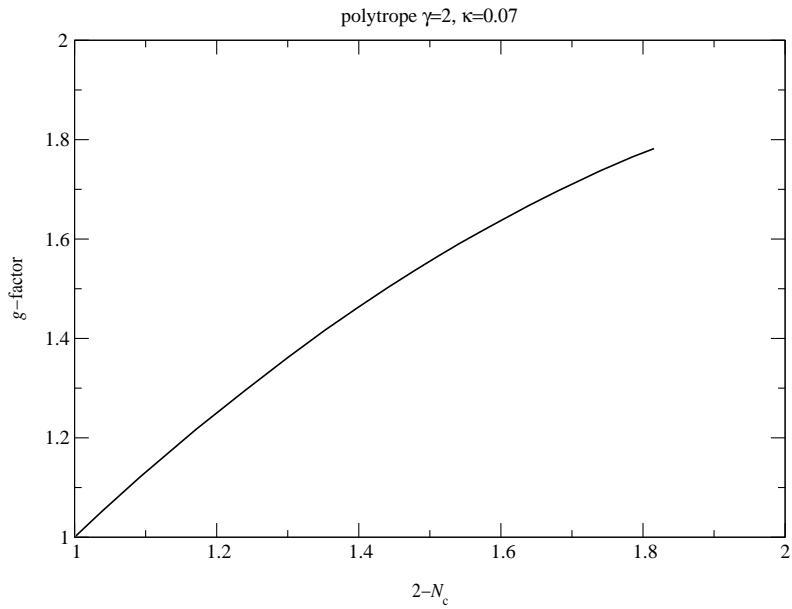}
\end{center}
\caption{Variation of the $g$-factor as a function of the central
lapse $N_c=N(r=0)$, for a polytrope. The angular velocity $\Omega =
10$ Hz is kept constant. }\label{f:polynu}
\end{figure}

The third parameter whose influence we want to study is the angular
velocity $\Omega$. Therefore, at fixed number of particles in the
star, and fixed total charge, we varied $\Omega$ from (almost) 0 to
the maximal value, called {\em Keplerian frequency} ($\Omega_{\rm K}$),
where the centrifugal force at the equator compensate the
gravitational attraction (shedding limit). The variation of the $g$-factor as a
function of $\Omega/\Omega_{\rm K}$, as well as that of a linear
combination of the central lapse $2 - N_c$ are displayed in
figure~\ref{f:polyfmb}. Both quantities show the same type of
behaviour: they are decreasing functions, mainly near $\Omega \sim
\Omega_{\rm K}$, but the overall change is relatively small, when
compared to that of figure~\ref{f:polymbnu}. We have explored here
high angular velocities, without any ``slow-rotation''
assumption, but the 
influence of these high velocities seems rather small. We have checked
at different masses, always obtaining the same kind of result. Here
again, we see that $2-N_c$ and $g$ follow the same type of evolution,
when varying $\Omega$. We therefore have plotted the $g$-factor, as a
function of $2-N_c$ as shown in figure~\ref{f:polynu}, when varying
the central value of the density $n_{\rm B}(r=0)$, like in
figure~\ref{f:polymbnu}. Contrary to that figure, there is no sign of 
the maximal mass point, the $g$-factor
being directly dependent on the strength of the gravitational field at
the centre of the star. It seems that the
gyromagnetic factor (\ref{e:def}) might be another indicator of the
strength of the gravitational field in self-gravitating objects: when
gravity is weak (well described by Newtonian theory), we have $g \sim
1$; when the star is very compact (even unstable) $g$ takes its
highest values. Finally, for a black hole, where gravity dominates
over other forces, we have $g=2$.

\subsection{Constant density models}\label{ss:incomp}

\begin{figure}
\begin{center}
\vspace{6mm}

\includegraphics[height=8cm]{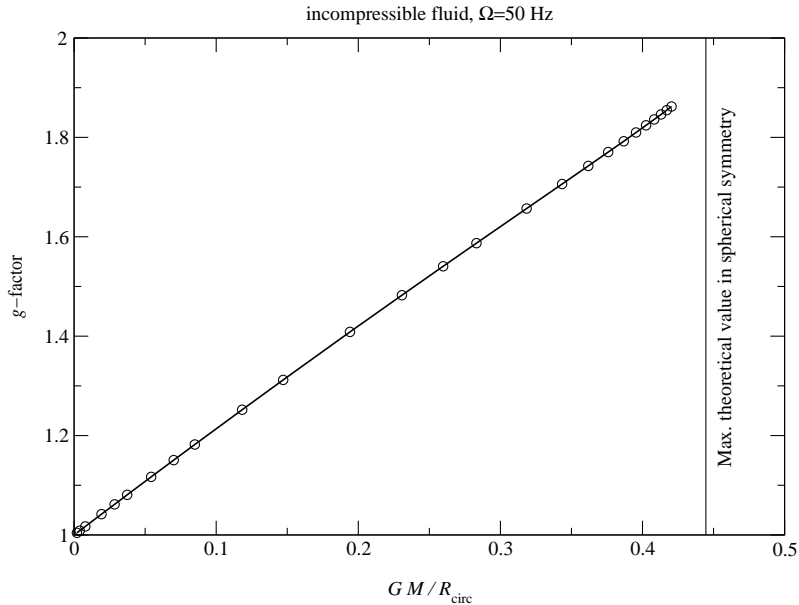}
\end{center}
\caption{Variation of the $g$-factor as a function of the compaction
parameter ${\cal C} = M/R_{\rm circ}$, at constant angular velocity
$\Omega = 50$ Hz, for an incompressible fluid. Circles show the model
that have been computed. The vertical line at ${\cal C} = 4/9$
corresponds to the maximal value of the compaction parameter for
spherically symmetric stars (see \cite{Buc59}).}\label{f:inccp} 
\end{figure}

In order to study the dependence of the results of the previous section on
the particular EOS, we first changed the values of $\gamma$ and
$\kappa$. The results obtained were qualitatively the same as in
previous section. Quantitatively, the trend described at the end of
previous section (figure~\ref{f:polynu} and the following discussion),
indicating that the $g$-factor be linked with the strength of the
gravitational field in the star was retrieved. We found that the more
compact the star, the higher the gyromagnetic factor. We checked this
result with several other equations of state, described in Salgado
{\it et al.\/} \cite{SalBGH94}, as well as with the ``strange quark
matter'' model (see Gourgoulhon {\it et al.\/} \cite{GouHLP99}), which
is giving very compact objects. Still, none of these EOS allowed for a
stable configuration with $g \gtrsim 1.8$. 

\begin{figure}
\begin{center}
\vspace{6mm}

\includegraphics[height=8cm]{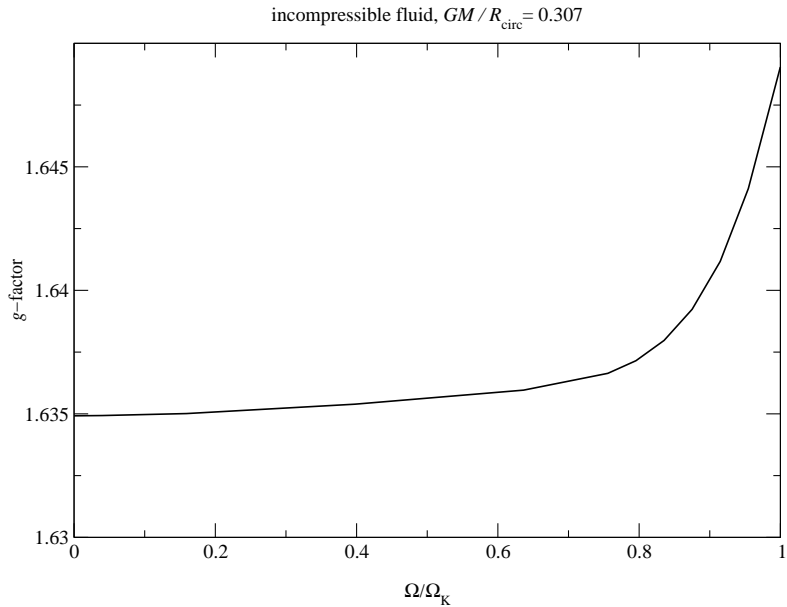}
\end{center}
\caption{Variation of the $g$-factor as a function of the angular 
velocity, at constant compaction parameter 
${\cal C} = M/R_{\rm circ} = 0.307$ for an incompressible
fluid. }\label{f:incfc} 
\end{figure}

Finally, we present here the limiting case of an incompressible fluid:
the EOS is such that $n_{\rm B}$ and $e$ are constant throughout the
star. Using this EOS in spherical symmetry, it can be shown (see,
\cite{Buc59}) that it gives an upper limit on the gravitational 
redshift at the surface of the star, when comparing with other EOS. We
here introduce a new dimensionless quantity called compaction parameter
\begin{equation}
{\cal C} = \frac{M}{R_{\rm circ}}, \label{e:defcp}
\end{equation}
with $R_{\rm circ}$ being defined by formula~(\ref{e:defrcirc}). In
Newtonian theory, this is the gravitational potential at the surface
of the star and ${\cal C} \ll 1$. On the other hand, for a
Schwarzschild black hole ${\cal C} = 0.5$ and, for relativistic stars
$0 \leq {\cal C} \leq 4/9$ (as derived by Buchdahl \cite{Buc59}). When
considering charged objects the maximal value is slightly increased, as
shown in Mak {\it et al.\/} \cite{MakDH01}. In
figure~\ref{f:inccp}, we display the dependence of the gyromagnetic
factor on this compaction parameter. As expected from previous
results, we find that $g$ goes to 1, when $\cal C$ becomes very small
(Newtonian limit). We could not reach the maximal theoretical value of
the compaction parameter for spherical stars (4/9), due to the
rotation and, perhaps, to the numerical algorithm (we look for a
solution by iteration, starting with a flat metric). Nevertheless, we
were able to get rather close to this limiting value, reaching the
highest gyromagnetic factors for self-gravitating fluids, with $g \sim
1.86$. The striking feature is that the $g$-factor seems to be
directly proportional to the compaction parameter $\cal C$ and, if one
extrapolates the line to the maximal value of $\cal C$, one finds that
the maximal value for the $g$-factor would be about 1.9, and
certainly well below 2. Using only low charge objects, this value of 2
is linked with the collapse limit that cannot be reached with our
stationary models.

It is now worth returning to the dependence of $g$ on the angular
velocity $\Omega$. If we fix the value of $\cal C$ for the star, how
does then gyromagnetic factor vary as a function of $\Omega$ ? The
answer is displayed 
in figure~\ref{f:incfc}, still in the case of an incompressible
fluid. We see that the dependence is very small, only about one
percent of variation between the non-rotating limit and the shedding
one. Results shown in figure~\ref{f:polyfmb} can have a new
interpretation: when $\Omega$ is increased, $\cal C$ decreases because
of centrifugal forces that go against gravity and therefore make
$R_{\rm circ}$ increase. Gravitational potential at the surface of the
star decreases because of the rotation, the same being true for the
gravitational field $N$ at the centre. The gyromagnetic factor
indicates that, at high angular velocities, the star is less
gravitationally bound. Comparing different models, equally bound, but
with different rotation rates (figure~\ref{f:incfc}), we see that the
effect of rotation is 
small, and must act on the increase of $g$ through the addition of
kinetic energy that contributes to the source terms of Einstein
equations. 

\subsection{Comparison with previous works}\label{ss:comp}

There have been some studies of the gyromagnetic factor in general
relativity, but none of them has considered physically consistent 
matter models. Much of interesting work has been done for slowly
rotating charged shells, starting with that of Cohen {\it et al.\/}
\cite{CohTW73}. The study that may be most closely related to ours
is that by Pfister and King \cite{PfiK02}, where the authors calculate
explicitly the gyromagnetic factor of a charged mass shell in slow
rotation approximation. The shell is infinitely thin and the authors
match two exact solutions of the Einstein equations in vacuum across
it. The properties of the energy-momentum tensor are then deduced from
the matching of both metrics. The advantage of their solutions is
that they were able to explore regimes with a very high charge and
compaction parameter. Unfortunately, it is difficult to compare
quantitatively their results with ours since one knows very little
about the properties of shell matter which, apart from
energy conditions, are not constrained. One might suppose that, in
general, these shells are not behaving like perfect
fluids. Qualitatively, both studies agree: for $Q/M \ll 1$ and taking
into account energy conditions, Pfister and King find that $g$ varies
between $\sim 1$, for a low compaction parameter, and 2 in the
collapse limit. With a similar kind of problem, Mustafa {\it et al.\/}
\cite{MusCP87} found that $g$ could reach values very close to 2, for
the charge-to-mass ratio less than unity and for the shell radius
approaching the event horizon value.

Garfinkle and Traschen \cite{GarT90} have calculated the $g$-factor of
a rotating massive loop of charged matter in the presence of a static charged
black hole. They have found that, for large radii of the loop, $g$
tended to 1, whereas they found $g\to 2$ for the radius approaching
the horizon. We retrieve (again qualitatively) the same results for
our self-gravitating and three-dimensional objects: a loop at spatial
infinity might be seen as undergoing a weak gravitational field, just like
self-gravitating body with a low compaction parameter. Let us also
mention here the very interesting work by Katz {\it et al.\/}, who
calculated the gyromagnetic ratio\footnote[1]{there is a factor 2
difference between their definition of $g$ and our (\ref{e:def})} in a
conformastationary metric \cite{KatBL99}. These axially symmetric
metrics can be seen as the external metrics for disc sources, made of
charged dust. They show that in those discs, hoop tensions are always
necessary to balance the centrifugal forces induced by the motion of
the rotating dust. The model therefore correspond to non-perfect
fluid, and they find $g=2$ (with our definition) for these
metrics.

\section{Conclusions}\label{s:conc}

We have studied the dependence of the gyromagnetic ratio
(\ref{e:def}) of self-gravitating rotating fluids on their mass
(number of particles), angular velocity and equation of state. We have
used a physical model in which we make the assumption that the fluid
is an insulator in uniform rotation, and we have neglected the
electromagnetic forces acting on the equilibrium of the fluid (low
charge approximation). These models have been solved 
numerically, with a code giving the solution in all space, which
enabled us to get the value of $g$ with a high accuracy (better than
$10^{-5}$) given by independent tests. We find that, with such
``stars'', $g$ can never
reach the value 2, characteristic of a charged rotating black
hole. The maximal value that can be achieved in our study is lower than
1.9. This gap may be linked with the fact that we have neglected
electromagnetic forces on hydrodynamic equilibrium and our study has
therefore been restricted to low charge-to-mass ratios. But it might
also be a result of that stationary relativistic stars, made of
perfect fluid, cannot reach values of the compaction parameter $M/R$
close to 1/2. In that sense, the $g$-factor is a good 
indicator of the strength of the gravitational field in an insulating
perfect fluid, but is little dependent on the angular velocity of the
star. In our study, the value $g=2$ seems linked only with the
black hole solution but, from other works, \cite{CohTW73}, \cite{PfiK02}
and \cite{MusCP87}, one can see that this may depend on the total
charge of the system. An important improvement of our 
work would be to allow for any charge of the system, that is
compatible with the stationarity assumption and therefore allow for
differential rotation, which may open new possibilities.

\ack We are grateful to H Pfister for suggesting this study and
stimulating discussions. We also thank J Bi\v{c}\'ak and P Peter who
gave us very helpful comments on the gyromagnetic ratio in relativity.


\section*{References}

\begin{thebibliography}{00}

\bibitem{PfiK03} Pfister H and King M 2003 \CQG {\bf 20} 205

\bibitem{BonGSM93} Bonazzola S, Gourgoulhon E, Salgado M and Marck J A
1993 {\it Astron. Astrophys.} {\bf 278} 421

\bibitem{BocBGN95} Bocquet M, Bonazzola S, Gourgoulhon E and Novak J
1995 {\it Astron. Astrophys.} {\bf 301} 757

\bibitem{CarPL01} Cardall C, Prakash M and Lattimer J 2001 {\it
Astrophys. J.} {\bf 554} 322

\bibitem{GraBGM01} Grancl\'ement P, Bonazzola S, Gourgoulhon E and
Marck J A 2001 {\it J. Comput. Phys.} {\bf 170} 231

\bibitem{Bona73} Bonazzola S 1973 {\it Astrophys. J.} {\bf 182} 335

\bibitem{BonG94} Bonazzola S and Gourgoulhon E 1994 \CQG {\bf 11} 1775

\bibitem{GouB94} Gourgoulhon E and Bonazzola S 1994 \CQG {\bf 11} 443

\bibitem{SalBGH94} Salgado M, Bonazzola S, Gourgoulhon E and Haensel
P 1994 {\it Astron. Astrophys.} {\bf 291} 155

\bibitem{GouHLP99} Gourgoulhon E, Haensel P, Livine R, Paluch E,
Bonazzola S and Marck J A 1999 {\it Astron. Astrophys.} {\bf 349} 851

\bibitem{MusCP87} Mustafa E, Cohen J M and Pechenick K R 1987 {\it
Int. J. Theor. Phys.} {\bf 26} 1189

\bibitem{Buc59} Buchdahl H A 1959 \PR {\bf 116} 1027

\bibitem{MakDH01} Mak M K, Dobson P N and Harko T 2001 {\it
Europhys. Lett.} {\bf 55} 310

\bibitem{CohTW73} Cohen J M, Tiomno J and Wald R M 1973 \PR D {\bf 7} 998

\bibitem{PfiK02} Pfister H and King M 2002 \PR D {\bf 65} 084033

\bibitem{GarT90} Garfinkle D and Traschen J 1990 \PR D {\bf 42} 419

\bibitem{KatBL99} Katz J, Bi\v{c}\'ak J and Lynden-Bell D 1999 \CQG
{\bf 16} 4023

\end{thebibliography}
\end{document}